\documentclass[twocolumn,showpacs,preprintnumbers,amsmath,amssymb]{revtex4}

\usepackage{graphicx}
\usepackage{dcolumn}
\usepackage{bm}

\begin{document}

\preprint{APS/123-QED}

\title{NMR Investigation of the Diluted Magnetic Semiconductor Li(Zn$_{1-x}$Mn$_x$)P (x = 0.1)}

\author{Cui Ding$^{1}$, Chuan Qin$^{1}$, Huiyuan Man$^{1}$, T. Imai$^{2,3}$} \author{F.L.
Ning$^{1,}$} \email{ningfl@zju.edu.cn}

\affiliation{$^{1}$Department of Physics, Zhejiang University,
Hangzhou 310027, China} \affiliation{$^{2}$Department of Physics and
Astronomy, McMaster University, Hamilton, Ontario L8S4M1,
Canada}\affiliation{$^{3}$Canadian Institute for Advanced Research,
Toronto, Ontario M5G1Z8, Canada}

\date{\today}


\begin{abstract}
We employ NMR techniques to investigate the nature of Mn spins in the I-II-V diluted magnetic semiconductor Li(Zn$_{1-x}$Mn$_x$)P (x = 0.1, Curie temperature $T_c$ = 25\ K).  We successfully identify the $^7$Li NMR signals arising from the Li sites adjacent to Mn$^{2+}$, and probe the static and dynamic properties of Mn spins.  From the NMR spin-lattice relaxation data, we show that the Mn spin-spin interactions extend over many unit cells.
\end{abstract}


\maketitle


The successful fabrication of (Ga$_{1-x}$Mn$_{x}$)As ferromagnetic
thin films by molecular beam epitaxy (MBE) techniques \cite{Ohno}
generated great interest in the research into diluted magnetic
semiconductors (DMS) \cite{MacDonald,Jungwirth,Dietl1}. The quality
of the epitaxial films steadily improved, and the Curie temperature,
$T_c$, has reached as high as 190\ K at the doping level of $x=12$ \%
\cite{Wang}. The spintronic applications of the DMS may soon become
a reality once the $T_c$ exceeds room temperature \cite{Zutic}.
Nonetheless, understanding the physical properties of the DMS
remains a major challenge \cite{Samarth,Chambers}. A fundamental
difficulty stems from the fact that the synthesis of the bulk form
of (Ga$_{1-x}$Mn$_{x}$)As is impossible beyond the very low
solubility limit of $x\sim$ 1\%, excluding the possibilities to
employ typical microscopic probes of magnetism, such as NMR and
neutron scattering techniques.

Recently, Deng et al. reported the successful synthesis of a bulk
Li(Zn$_{1-x}$Mn$_x$)As DMS with $T_c$ as high as $\sim$ 50 K based
on the I-II-V direct-gap semiconductor LiZnAs \cite{Deng}. The
structure of LiZnAs, shown in Fig. 1(a), can be viewed as an
analogue of  GaAs with the Zinc-blende structure, see Fig. 1(b);
notice that if we replace Ga$^{3+}$ sites of GaAs with Zn$^{2+}$ and
place Li$^{+}$ ions between Zn$^{2+}$, we obtain LiZnAs. Also notice
that the total number of electrons in (ZnAs)$^{-}$ is equal to that
of GaAs.  LiZnAs has a direct band gap of $\sim$ 1.6 eV
\cite{Bacewicz,Kuriyama1, Kuriyama2,Wei}, which is comparable to
that of GaAs (1.42 eV). On the other hand, the isovalent substitution of Mn$^{2+}$
for Zn$^{2+}$ resembles that of II-VI DMS, although it is difficult to
control the carrier density in the latter \cite{Furdyna,Shand}.
Recent $\mu$SR measurements of
Li(Zn$_{1-x}$Mn$_x$)As \cite{Deng} have established that the
magnetically ordered volume reaches 100 $\%$ below $T_c$, and the
magnitudes of the ferromagnetic exchange coupling and the ordered
moment are comparable to those of (Ga,Mn)As \cite{Dunsiger}. More
recently, some of us \cite{Deng2} found that the iso-structural
direct-gap semiconductor LiZnP \cite{Bacewicz,Kuriyama} also
undergoes a ferromagnetic transition upon Mn doping, and its bulk
magnetic properties are very similar to those of LiZnAs.

\begin{figure}[b] \centering
\includegraphics[width=3in]{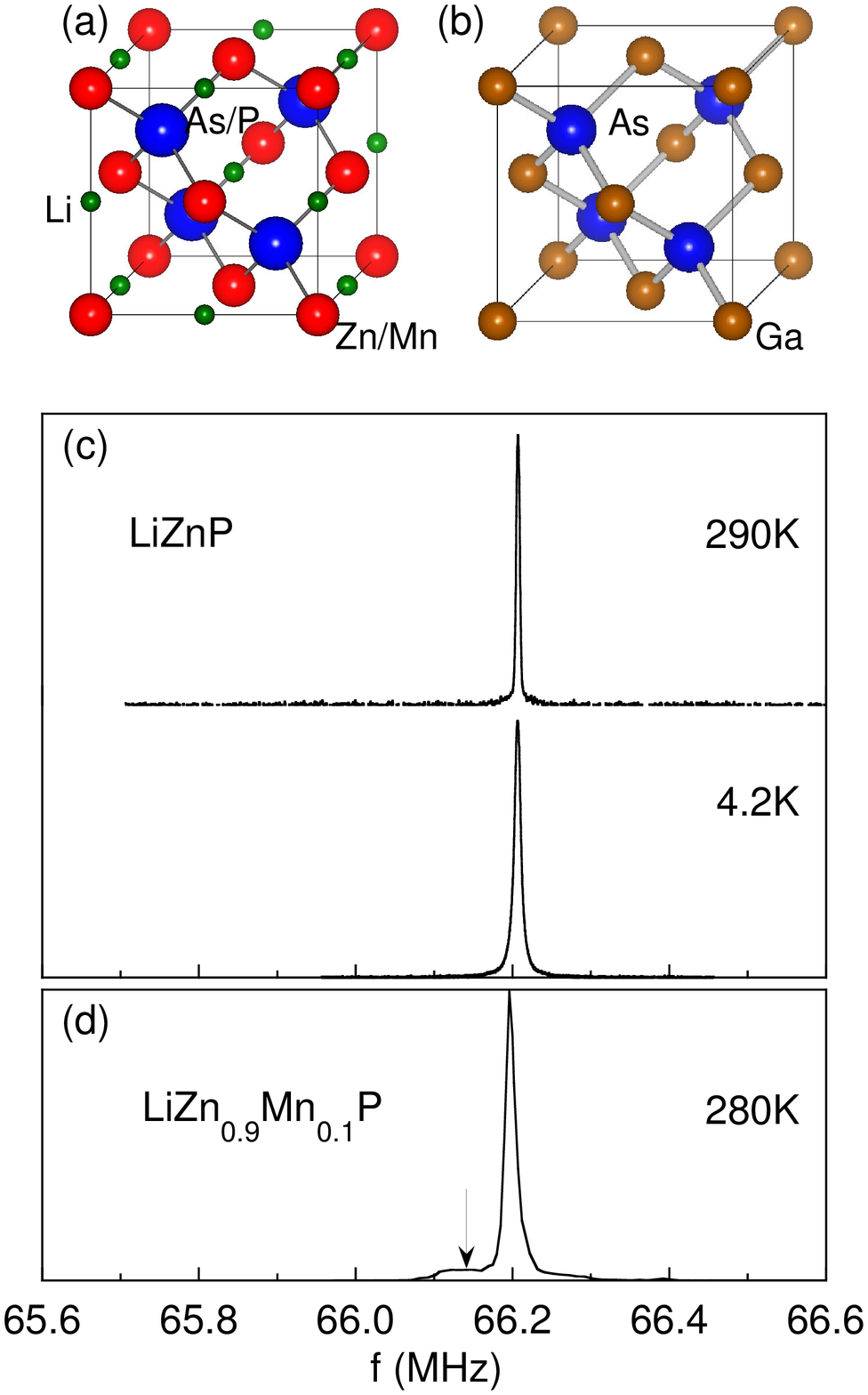} 
\caption{\label{Fig1:epsart} (Color online) The crystal structure of
(a) LiZnP, and (b) GaAs. The representative $^7$Li lineshapes in 4
Tesla of (c) the parent compound LiZnP, and (d)
LiZn$_{0.9}$Mn$_{0.1}$P. The vertical arrow marks Li(Mn), the n. n. site
of doped Mn.}
\end{figure}

In this paper, we use $^7$Li NMR to investigate
LiZn$_{0.9}$Mn$_{0.1}$P ($T_c$ = 25 K). Since NMR probes the
electronic properties in the immediate vicinity of the observed
nuclear spins, our NMR measurements will provide {\it local} and
{\it site-selective} information for LiZn$_{0.9}$Mn$_{0.1}$P. In this
context, it is worth recalling that NMR measurements provided vital
microscopic information on the nature of dilute magnetic spins doped
into, e.g. simple metals \cite{Boyce} and high $T_c$ superconductors
\cite{Bobroff}. By successfully identifying $^7$Li NMR signals
arising from the nearest-neighbor (n. n.) Li site of doped Mn
(denoted as the Li(Mn) site hereafter), we were able to probe the
static and dynamic properties of Mn magnetic moments using NMR in
the paramagnetic state of DMS for the first time.

The polycrystalline LiZnP and LiZn$_{0.9}$Mn$_{0.1}$P were
synthesized by the solid state reaction method. High purity elements
of Li (99.9\%), Zn (99.9\%), Mn (99.99\%), and P (99\%) were mixed
and slowly heated to 450$^{\circ}$C in evacuated silica tubes, and
held for 48 hours before cooling down to room temperature at the
rate of 20$^{\circ}$C/h. X-ray diffraction showed that both
specimens are single phase with a cubic structure. The lattice
constants are 5.7371 {\AA} and 5.7434 {\AA} for LiZnP and
LiZn$_{0.9}$Mn$_{0.1}$P, respectively. We also confirmed from the
$dc$-magnetization and $\mu$SR measurements that the ferromagnetic
ordering is of bulk nature with $T_c$ = 25 K. The Hall effect
measurements conducted for samples prepared under the identical
conditions showed that LiZn$_{0.9}$Mn$_{0.1}$P is a hole-doped
semiconductor with a very low carrier concentration of $p \sim 3
\times 10^{17}$ cm$^{-3}$ \cite{Deng2}. We will report the complete
details of synthesis and characterization of the
LiZn$_{0.9}$Mn$_{0.1}$P sample by X-ray diffraction, electrical
resistance, and $\mu$SR elsewhere \cite{Deng2}.

\begin{figure}[!b] \centering
\centering
\includegraphics[width=3in]{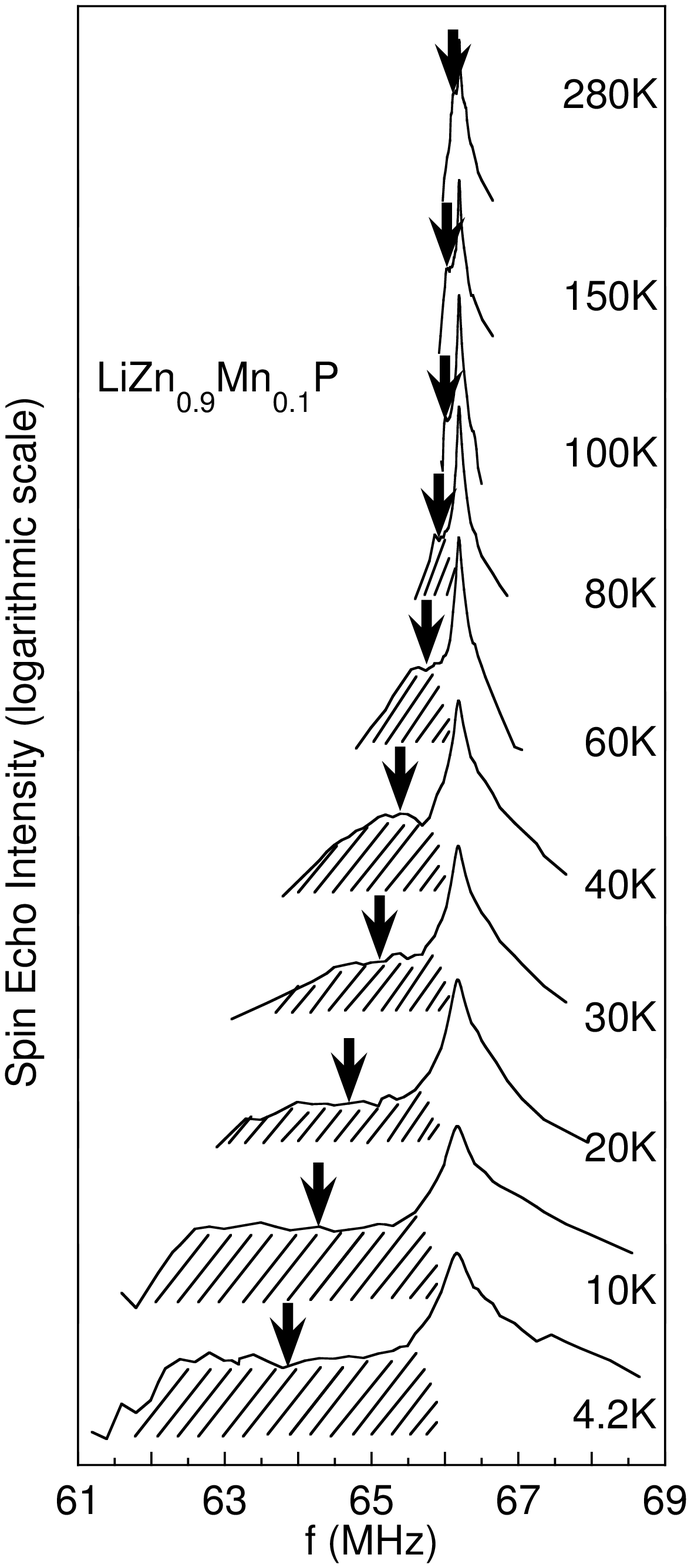}
\caption{\label{Fig2:epsart} The temperature dependence of
the $^7$Li NMR line of Li(Zn$_{0.9}$Mn$_{0.1}$)P plotted in a
semi-log scale. Shadowed areas mark Li(Mn), and vertical arrows mark
the position where we measured $\frac{1}{T_1}$ of the Li(Mn) site.
Note that the total spin echo intensity is not normalized, and the
origin is shifted vertically for different temperatures for
clarity.}
\end{figure}

We show representative $^7$Li NMR lineshapes of the parent compound
LiZnP in Fig.\ 1 (c). $^7$Li has nuclear spin $I = \frac{3}{2}$, and the
gyromagnetic ratio is $^7\gamma_n/2\pi$ = 16.546 MHz/Tesla. In the
external magnetic field $B_{ext}$ = 4 Tesla, resonance would
take place at $f_0$ = $(^7\gamma_n/2\pi)B_{ext}$ = 66.184 MHz in the
absence of hyperfine interactions with electrons. In general,
three NMR peaks for the $I_z$ = m to m+1 transition (m =
$-\frac{3}{2}$, $-\frac{1}{2}$, $\frac{1}{2}$) could arise for each
inequivalent Li site, if there exists significant interaction
between the nuclear quardrupole moment and the electric field
gradient (EFG). For the parent compound LiZnP only one Li peak is
observed at $f_0$ $\sim$ 66.206 MHz, as shown in Fig. 1(c). This is
because Li atoms are located in a nearly cubic environment, and the
EFG is extremely small due to the high symmetry of the lattice. The half height
full width (HHFW) of this peak is $\sim$ 5 KHz at 290 K, indicating
the high quality of the polycrystalline sample. The very small NMR
Knight shift (a.k.a. the paramagnetic frequency shift) \cite{Slichter},
$^{7}K = (66.206 - 66.184)/66.184 = 0.0003$ (i.e. +0.03 \%, or +300
ppm), is consistent with the absence of conduction electrons or
magnetic moments.

\begin{figure}[!b]
\centering\vspace*{-1.5cm}
\includegraphics[width=3.3in]{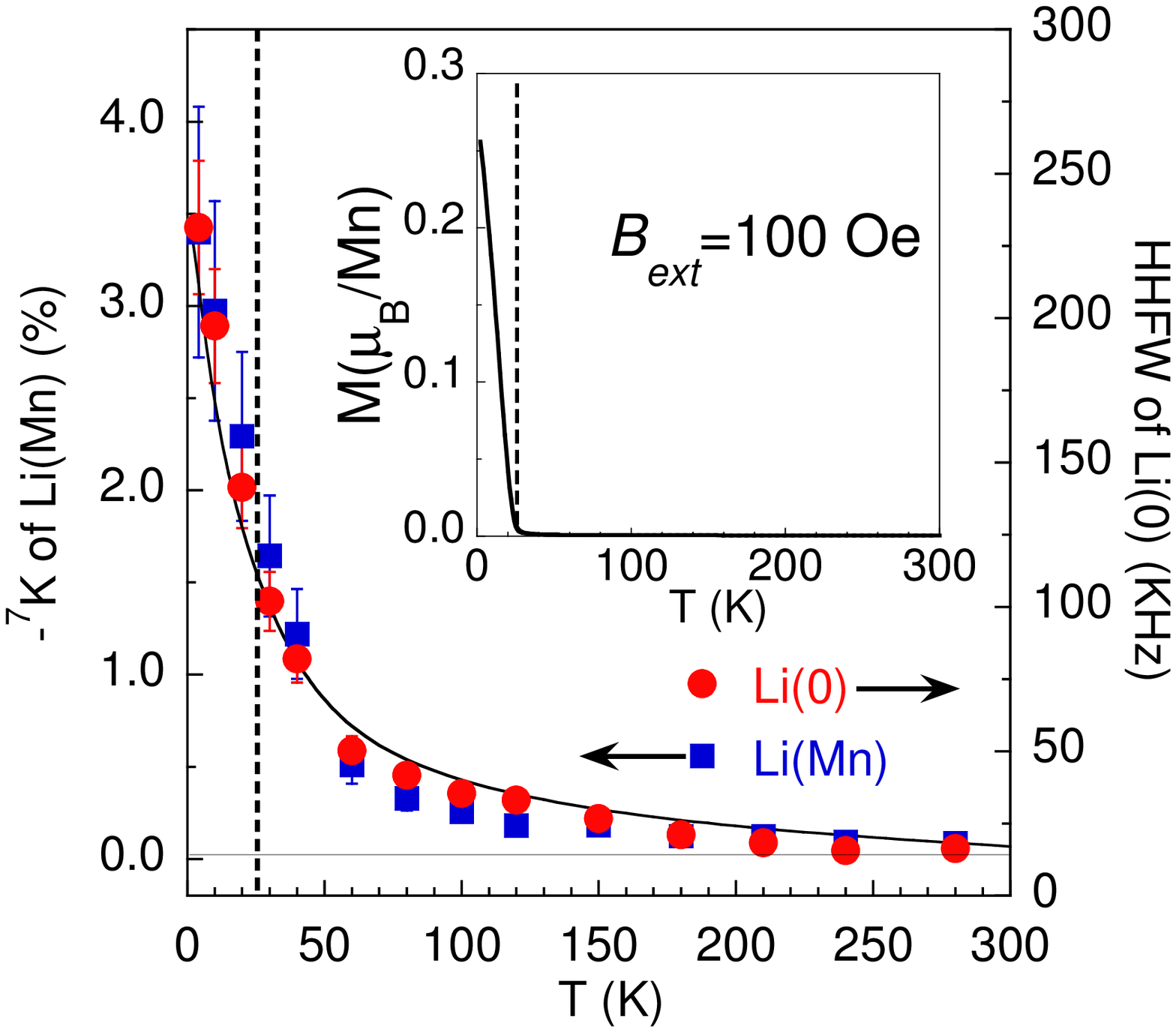} \vspace*{+1.5cm}\\
\caption{\label{Fig3:epsart}(Color online) The temperature
dependence of the $^{7}$Li NMR Knight shifts, -$^{7}K$, at the
Li(Mn) sites ($\blacksquare$).  The sign of $^{7}K$ is reversed, to
take into account the negative sign of the hyperfine coupling
$A_{o}$. The horizontal solid line represents $^{7}K_{orb}$ =
0.05\%. Also shown is the HHFW of Li(0) in Li(Zn$_{0.9}$Mn$_{0.1}$)P
($\bullet$). For comparison, we overlay the temperature dependence
of the bulk $dc$-magnetization $M$ at 4 Tesla measured with SQUID
(solid curve, normalized at 4.2 K). Inset: the $dc$-magnetization
$M$ measured at $B_{ext}$ = 100 Oe; $M$ increases sharply below
$T_c$ = 25 K. Notice that this feature is suppressed at $B_{ext}$ =
4 Tesla. Dashed lines mark $T_c$ = 25 K.}
\end{figure}

The substitution of Mn drastically alters the $^7$Li NMR lineshape
of LiZn$_{0.9}$Mn$_{0.1}$P as shown in Fig. 1(d). To enhance the
extra features, we plot the temperature dependence of $^7$Li NMR
lineshapes in a semi-logarithmic scale in Fig. 2. In addition to the
relatively sharp but somewhat asymmetrical Li(0) peak that
corresponds to the single peak of LiZnP, a broad hump appears on the
lower frequency side, as marked by a downward arrow in Fig. 1(d) and
the shadowed area in Fig. 2. This broad hump should be attributed to
Li(Mn) sites whose magnetic environment is altered by the substitution of the Zn$^{2+}$ sites in their vicinity by Mn$^{2+}$ ions. Upon decreasing
the temperature from 280 K to 30  K, the Li(Mn) peak shifts
progressively towards lower frequencies, and becomes as broad as 3
MHz. These striking featues of the Li(Mn) peak should be attributed
to the spin polarization transferred from the n. n. Mn sites through
weak but finite hybridizations between Mn $3d$ and Li orbitals. We
define the NMR Knight shift of the Li(Mn) sites at the center of the
broad hump, as marked by downward arrows in Fig. 2. We summarize the
temperature dependence of $-^{7}K$ measured in $B_{ext} = 4$\ T for
the Li(Mn) sites in Fig. 3. $-^{7}K$ exhibits identical temperature
dependence to the bulk magnetization $M$ as measured by SQUID in
$B_{ext} = 4$\ T (solid curve). Also presented in the inset is $M$ measured in
$B_{ext} = 0.01$\ T, establishing the bulk ferromagnetism below
$T_{c}=25$\ K. In general,
$^7K=^7K_{spin}+^7K_{orb}=(A_{0}/g\mu_{B})\chi_{spin}+^7K_{orb}$.
The small orbital shift $^7K_{orb}$ is temperature independent
($\sim +0.05 \%$ in the present case), while the hyperfine coupling between the Mn electron spins and the Li
nuclear spin, $A_{0} = -12$\ kOe/${\mu}_{B}$, turns out to be negative for the Li(Mn) peak, hence the
overall sign of $^7K < 0$. Note that $-^7K$ reaches as large as 3.5
\%.

In contrast with the Li(Mn) sites, $^7K$ at Li(0) remains small and
comparable to that in nonmagnetic LiZnP (see Fig. 1). The HHFW of
Li(0), however, is much broader than that of LiZnP ($\sim$5 KHz),
and continuously increases from 20 KHz at 280 K to 240 KHz at 4.2 K.
The fact that HHFW of Li(0)  is an order of magnitude broader in
LiZn$_{0.9}$Mn$_{0.1}$P suggestss that the hyperfine magnetic fields
from the doped Mn strongly affect {\it all} Li sites, including
those without Mn atoms at n.n. Zn sites. Our findings in Fig. 3 are
consistent with our $\mu$SR results observed for the same sample
that the ferromagnetic volume fraction reaches 100 $\%$
\cite{Deng2}.

\begin{figure}[!b] \centering
\centering
\includegraphics[width=3in]{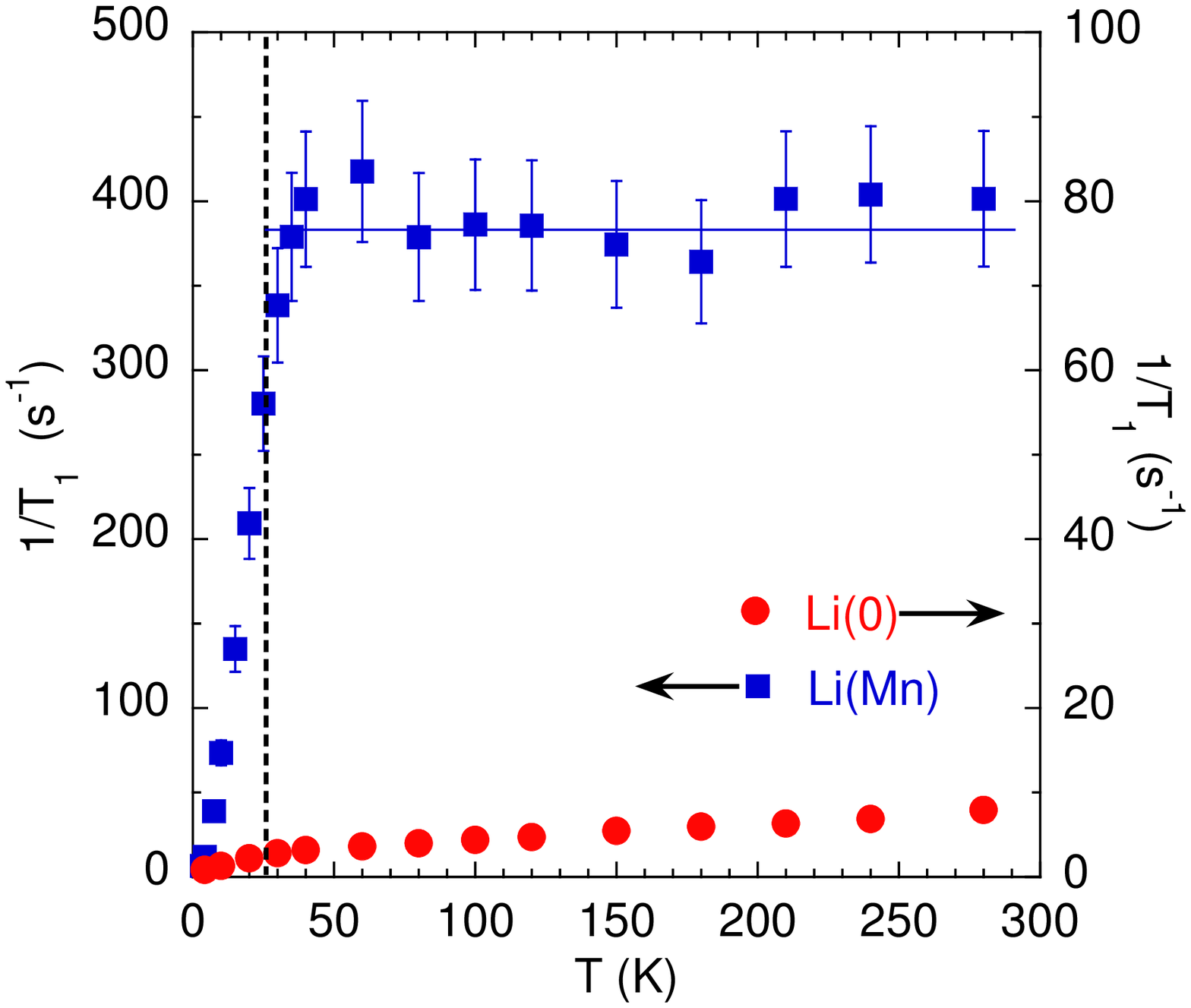}
\caption{\label{Fig4:epsart} (Color online) $\frac{1}{T_{1}}$ of
Li(0)($\bullet$) and Li(Mn)($\blacksquare$) of
LiZn$_{0.9}$Mn$_{0.1}$P, with the solid line as a guide for the eyes. The dashed line
marks $T_c$ = 25 K.}
\end{figure}

To gain additional insight into the nature of Mn spins, we also
measured the nuclear spin-lattice relaxation rate $\frac{1}{T_{1}}$
at the Li(0) peak and the center of the broad Li(Mn) peak, as marked
by downward arrows in Fig. 2. We note that we do not observe
satellite peaks corresponding to the $\pm\frac{1}{2}$ -
$\pm\frac{3}{2}$ transitions for both Li(0) and Li(Mn) sites.  This
is not surprising, in view of the fact that even the Li(Mn) sites
are still located in a high symmetry environment; Mn$^{2+}$ ions are
isovalent with Zn$^{2+}$ ions, and the difference between the ionic
radius of Mn$^{2+}$ (0.083 nm) and that of Zn$^{2+}$ (0.074 nm) is
small.  We also confirmed that the radio frequency pulse width
optimized for the Li(Mn) and Li(0) peaks is identical with that of
LiZnP, implying that the Li(0) peak does not arise from the
$+\frac{1}{2}$ to $-\frac{1}{2}$ transition alone.  If the sharper
Li(0) peak in the middle originated only from the $+\frac{1}{2}$ to
$-\frac{1}{2}$ transition, the pulse width should be narrower by a
factor of 2.  These findings indicate that we are observing a
superposition of the $+\frac{1}{2}$ to $-\frac{1}{2}$ and
$\pm\frac{1}{2}$ to $\pm\frac{3}{2}$ transitions in all cases, which
would lead to a single exponential recovery curve for the
spin-lattice relaxation process. Since our sample is an alloy-doped
with Mn$^{2+}$ magnetic moments, however, $\frac{1}{T_{1}}$ is bound
to have a distribution.  We therefore use a stretched exponential
function to fit the $T_{1}$ recovery data in all cases (see
Supplemental Materials for the measurement procedures). This is a
standard practice for NMR studies of inherently disordered magnetic
materials.

We summarize the results of $\frac{1}{T_{1}}$ in Fig. 4. The nuclear
spin-lattice relaxation time $T_{1}$ represents the time scale
during which Li nuclear spins relax to their thermal equilibrium
after the absorption of radio frequency pulses. $\frac{1}{T_{1}}$
becomes larger when more low energy spin excitations associated with
the Mn spins or the conduction electrons exist. Theoretically, the
spin contribution to $\frac{1}{T_{1}}$ may be written using the
imaginary part of the dynamical electron spin susceptibility
${\chi''({\bf q},f_{o})}$ as $\frac{1}{T_{1}} \propto T \sum_{{\bf
q}} | A({\bf q}) |^{2} \frac{\chi''({\bf q},f_{o})}{f_{o}}$
\cite{Moriya}, where $A({\bf q})$ is the hyperfine form factor
\cite{Moriya}. In the parent compound LiZnP, $\frac{1}{T_{1}}$ of
the Li(0) sites is as slow as $\sim$ 0.004 $s^{-1}$. Such very slow
relaxation rates are typical for non-magnetic insulators, and
consistent with the fact that both Mn moments and conduction
carriers are absent in the non-magnetic semiconductor LiZnP.

Upon Mn doping, the relaxation rate increases dramatically.
$\frac{1}{T_{1}} \sim 400$ s$^{-1}$ at the Li(Mn) site observed
above $T_{c}$ is five orders of magnitude enhanced compared with the
non-magnetic LiZnP. This finding provides evidence for the presence of
strong, low-frequency Mn spin fluctuations in the paramagnetic state
above $T_{c}$. The suppression of $\frac{1}{T_{1}}$ below $\sim
T_{c}$ indicates that these spin fluctuations are suppressed once
ferromagnetic long-range order is established. The temperature
independence of $\frac{1}{T_{1}}$ in the broad temperature range
from 280\ K down to $\sim 30$\ K also implies that the Mn spin-spin
correlations are in the so-called exchange narrowed regime
\cite{Moriya1956}, where the energy scale of the thermal disturbance
($k_{B}T$) is comparable to or greater than that of the typical Mn-Mn
spin interaction $| J |$.  Using the Gaussian approximation for the spin-spin correlation function, we can express $\frac{1}{T_{1}} = \sqrt{2\pi}
\frac{S(S+1)}{3 \omega_{e}}(\frac{A_{0}}{\hbar})^{2} \sim 400
s^{-1}$ \cite{Moriya1956}, where $S\lesssim 5/2$ \cite{Deng} and
$A_{0} =-12$ kOe/$\mu_{B}$ as determined from Fig.
3.  Therefore we estimate the characteristic frequency of the Mn spin
fluctuations as $\omega_{e} \sim 1.1 \times 10^{15}$\ rad/s in the
paramagnetic state above $T_{c}$ .

This large value of $\omega_{e}$ sets a strong constraint on the
nature of Mn-Mn spin interactions.  First,
we recall that $\omega_{e}^{2} = \frac{2}{3} z S(S+1)
(\frac{J}{\hbar})^{2}$ within the Gaussian approximation \cite{Moriya1956}, where $z$  is
the number of Mn sites within the range of Mn-Mn interactions.  Since
the high temperature regime with a constant $1/T_{1} \sim 400$
sec$^{-1}$ extends to $\sim T_{c}$, the typical Mn-Mn spin
interaction energy scale $|J|$ is at most $\sim 100$ K.  This
implies that the number of Mn sites within the range of Mn-Mn
interactions is as large as $z \sim 10^{3}$.  While our estimation
may be somewhat crude, our $1/T_{1}$ data establishes that each Mn
spin interacts with a very large number of other Mn spins, rather
than just a few Mn sites in the immediate vicinity.  Our finding is consistent with
the earlier theoretical proposal that the Mn-Mn spin interactions
are caused primarily by the long-range interactions, such as the
p-d Zener exchange interaction mediated by conduction carriers
(doped holes or electrons) \cite{Dietl}, which is the semiconductor
analogue of the RKKY interaction in metals.  We note, however, that
our finding based on $\omega_{e}$ only implies the extended nature
of the Mn-Mn interactions, and does not rule out other  interaction
mechanisms.

$\frac{1}{T_{1}}$ at the main Li(0) peak of LiZn$_{0.9}$Mn$_{0.1}$P
is slower than the Li(Mn) site by a factor of $\sim 50$. This is
consistent with the fact that Li(0) sites do not have Mn at their n.
n. sites, hence the effects of Mn spin fluctuations are much more
limited due to weaker hyperfine couplings. Nonetheless,
$\frac{1}{T_{1}}$ at the Li(0) peak also exhibits a kink around
$T_{c}$, indicating that Li(0) sites are indeed under the influence
of ferromagnetic Mn spin fluctuations as well. The observed linear
increase of $\frac{1}{T_{1}} \sim a + bT$ above $T_{c}$ suggests
that Li(0) nuclear spins may have two separate relaxation
mechanisms.  The constant $a$-term is caused by Mn spin fluctuations,
in analogy with the constant behavior observed for Li(Mn), while the small T-linear $bT$-term may be attributed to the Korringa process
\cite{Slichter} arising from the Fermi surface excitations of a
small number of conduction carriers. These carriers  might also
facilitate the interactions between distant Mn sites.

To summarize, the successful synthesis of the bulk form of I-II-V DMS,
LiZn$_{0.9}$Mn$_{0.1}$P with $T_{c} = 25$ K, enabled us to
investigate the nature of its ferromagnetism by microscopic $^{7}$Li
NMR techniques.  In particular, by successfully identifying the
$^{7}$Li NMR signals arising from the Li sites n. n. to the Mn
spins, we were able to probe the static and dynamic magnetic
susceptibility of Mn spins through the NMR Knight shift and
$1/T_{1}$.  We deduced the characteristic
Mn  spin fluctuation frequency $\omega_{e}$, and showed that the
Mn-Mn spin interaction extends many unit cells.  Regardless of the
exact mechanism of such interactions, our findings explain why DMS
could exhibit a relatively high $T_{c}$ with a low density of Mn.

The work at Zhejiang University was supported by National Basic
Research Program of China (No.2011CBA00103), NSF of China
(No.11274268), Zhejiang Provincial Natural Science Foundation of
China (LY12A04006) and Fundamental Research Funds for Central
Universities (2013QNA3016). The work at McMaster was supported by
NSERC and CIFAR. F.L. Ning acknowledges helpful discussions with
Y.J. Uemura, C.Q. Jin, F.C. Zhang, X. Wan and C. Cao.
\\



\end{document}